\documentclass{article}

\usepackage{arxiv}
\usepackage{amsmath}
\usepackage[utf8]{inputenc} 
\usepackage[T1]{fontenc}    
\usepackage{hyperref}       
\usepackage{url}            
\usepackage{booktabs}       
\usepackage{amsfonts}       
\usepackage{nicefrac}       
\usepackage{microtype}      
\usepackage{lipsum}		
\usepackage{graphicx}
\usepackage[square,numbers]{natbib}
\usepackage{doi}
\usepackage{xcolor}
\usepackage[section]{placeins}

\usepackage{color}
\newcommand\revs[1]{\textcolor{black}{#1}}
\newcommand\nr[1]{\textcolor{black}{#1}}

\title{Locality sensitive hashing via mechanical behavior}

	\author{ \href{https://orcid.org/0000-0001-8099-3468}{\includegraphics[scale=0.06]{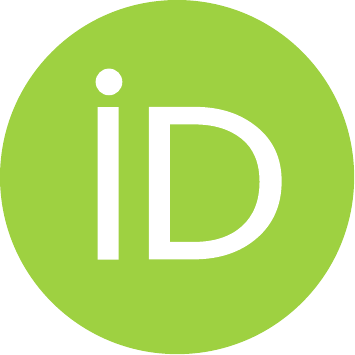}\hspace{1mm}Emma~Lejeune$^1$} \\
	Department of Mechanical Engineering\\
	Boston University\\
	Boston, MA \\
	\texttt{elejeune@bu.edu} \\
 	\And
        \href{https://orcid.org/0009-0000-3325-1410}{\includegraphics[scale=0.06]{orcid.pdf}\hspace{1mm}Peerasait~Prachaseree}\\
	Department of Mechanical Engineering\\
	Boston University\\
	Boston, MA 02215 \\
	\texttt{pprachas@bu.edu} \\
}

\renewcommand{\shorttitle}{Locality sensitive hashing via mechanical behavior}

\hypersetup{
pdftitle={Locality sensitive hashing via mechanical behavior},
pdfauthor={Peerasait~Prachaseree,Emma~Lejeune},
pdfkeywords={physical computing, morphological computing, programmable matter, mechanical hashing},
}

\begin{document}
\maketitle

\footnotetext[1]{corresponding author}

\begin{abstract}

From healing wounds to maintaining homeostasis in cyclically loaded tissue, living systems have a phenomenal ability to sense, store, and respond to mechanical stimuli. Broadly speaking, there is significant interest in designing engineered systems to recapitulate this incredible functionality. In engineered systems, we have seen significant recent computationally driven advances in sensing and control. And, there has been a growing interest --- inspired in part by the incredible distributed and emergent functionality observed in the natural world --- in exploring the ability of engineered systems to perform computation through mechanisms that are fundamentally driven by physical laws. In this work, we focus on a small segment of this broad and evolving field: \textit{locality sensitive hashing via mechanical behavior}. Specifically, we will address the question: can mechanical information (i.e., loads) be transformed by mechanical systems (i.e., converted into sensor readouts) such that the mechanical system meets the requirements for a locality sensitive hash function?  Overall, we not only find that mechanical systems are able to perform this function, but also that different mechanical systems vary widely in their efficacy at this task. Looking forward, we view this work as a starting point for significant future investigation into the design and optimization of mechanical systems for conveying mechanical information for downstream computing.

\end{abstract}


\section{Introduction}
\label{sec:intro}

 From the cells embedded in our skin deciding if they should activate to heal a wound \citep{das2021extracellular,sree2020computational}, to robotic systems dexterously manipulating delicate objects \citep{amadeo2022soft,gerald2022soft,yang2021grasping}, the ability to effectively transmit and interpret mechanical signals can lead to incredible functionality \citep{chin2019automated,truby2019soft,spielberg2021co}. 
In natural systems, this ability leads to complex emergent behavior such as the maintenance of homeostasis in mechanically loaded tissue \citep{meador2020tricuspid,han2022bayesian}. 
And in engineered systems we can design for responsiveness by controlling the transmission of mechanical signals through material selection and structural form \citep{osborn2018prosthesis,truby2021designing}.

Transmission and interpretation of mechanical signals is especially relevant to growing interests in ``morphological computing'' \citep{fuchslin2013morphological}, ``physical learning'' \citep{stern2022learning}, and ``programmable matter'' \citep{hawkes2010programmable,chen2021reprogrammable}. Broadly speaking, these are all paradigms where a physical system is either programmed, or used to perform some form of ``computation.'' For example, researchers have experimentally realized physical logic gates \citep{song2019additively,el2021digital,meng2021bistability}, as well as responsive mechanisms that trigger functional behavior when activated \citep{chen2018harnessing,zhu2020elastically,wei2022temperature}. And, within the scope of dynamical systems, researchers have used physical bodies to perform ``reservoir computing'' where a higher dimensional computational space is created by multiple non-linear responses to an input signal \citep{hauser2021physical,nakajima2018exploiting}, and cryptographic hashing where researchers have shown that chaotic hydrodynamics can be used to store and manipulate information in a fluid system \citep{gilpin2018cryptographic}. In this paper, we will focus on a small segment of this broad and emerging field: \textit{locality sensitive hashing via mechanical behavior}. Here, our goal is to explore this specific type of computation in the context of mechanical systems. 

\begin{figure}[h!]
	\centering
	\includegraphics[width=0.45\textwidth]{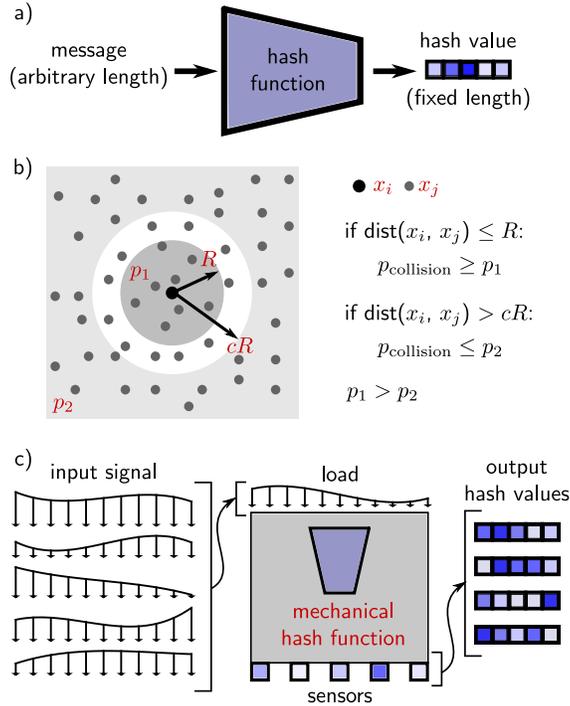}
	\caption{a) Schematic illustration of a generally defined hash function; b) Schematic illustration of the requirements for locality sensitive hashing where $p_{\mathrm{collision}}$ is the probability of a hash collision that is larger for inputs that are closer together; c) Schematic illustration of a mechanical system performing locality sensitive hashing.}
	\label{fig:fig1}
\end{figure}

Hashing, the process of converting arbitrarily sized inputs to outputs of a fixed size, is schematically illustrated in Fig. \ref{fig:fig1}a \citep{buchmann2004introduction}.  
In most popular applications of hashing (e.g., storing sensitive information), it is desirable to minimize collisions (i.e., the occurrence of different inputs mapping to the same output) and obfuscate the relationship between inputs and outputs.
However, there has been a growing interest in alternative types of hashing algorithms -- specifically hashing algorithms for applications such as similarity search, see Fig. \ref{fig:fig1}b \citep{wang2014hashing}. 
In these algorithms, the goal is to compress input data while preserving essential aspects of the relationship between input data points. 
In Section \ref{sec:meth_lsh}, we lay out the mathematical definition for ``locality sensitive hashing'' \citep{jafari2021survey,pauleve2010locality}. 
In this paper, we will focus on the concept schematically illustrated in Fig. \ref{fig:fig1}c. Can mechanical information (i.e., loads) be transformed by mechanical systems (i.e., converted into sensor readouts) such that the mechanical system meets the requirements for a locality sensitive hash function?  
In exploring this specific type of computing in mechanical systems, our goal is to lay a solid ideological foundation for future applications of physical computing where mechanical systems are tailored to act as a ``physical computing layer'' that transforms mechanical information to enable downstream responsiveness and control.

The remainder of the paper is organized as follows. In Section \ref{sec:methods}, we further define locality sensitive hashing, elaborate on the concept of mechanical systems as locality sensitive hash functions, and define an example problem to explore the performance of different mechanical systems for locality sensitive hashing. 
Then, in Section \ref{sec:res_disc}, we show the results of our investigation of our example problem, and conclude in Section \ref{sec:conclusion}.
Overall, our goal is threefold: (1) to introduce the concept of locality sensitive hashing in the context of mechanical systems, (2) to provide a straightforward ``proof of concept'' that mechanical systems can be used to perform locality sensitive hashing, and (3) to lay the foundation for future investigations on \revs{optimizing} mechanical behavior to perform hashing for similarity search.

\section{Methods}
\label{sec:methods}

We will begin in Section \ref{sec:meth_lsh} by defining Locality Sensitive Hashing (LSH), then in Section \ref{sec:methods_ss} we will demonstrate how the concept behind LSH can be applied to mechanical systems as a ``proof of concept.'' Finally, in Section \ref{sec:example_problem} we define an example problem that will set up the main investigation presented in this paper.

\subsection{Locality sensitive hashing}
\label{sec:meth_lsh}

\revs{In simple terms, a ``hash function'' is a function that maps input data of an arbitrary size to a fixed size output, referred to as a ``hash value''  \citep{chi2017hashing}. This is schematically illustrated in Fig. \ref{fig:fig1}a.
Hash functions have broad societal applications ranging from storing passwords, to checking if files match, to enabling data structures (e.g., dictionaries in Python)
\citep{python}.
For these applications, hash functions are designed to minimize ``hash collisons.'' 
To minimize ``hash collisons,'' hash functions typically convert similar yet different inputs to drastically different hash values \citep{rivest1992md5}.} Therefore, for typical hash algorithms, it would not make sense to perform downstream applications that rely on the distance between hash values. However, there has been recent interest in an alternative type of hash algorithm referred to as ``locality sensitive hashing'' where the goal is to create hash functions that \textit{encourage} collisions between similar inputs \citep{wang2014hashing}. For these Locality Sensitive Hash (LSH) approaches, similar inputs should lead to similar or identical hash values. To date, these techniques have primarily  been used for dimensionality reduction prior to nearest neighbor search \citep{slaney2008locality}. More formally, we can introduce LSH through the following definition \citep{wang2014hashing}. 

First, we describe our input data as points in a $N$ dimensional metric space $\mathcal{M}$ with distance function $d$. Here we will choose $d$ as the $L_{\infty}$ norm\footnote{\revs{Alternative choices of $||L||$ would also be acceptable, here we choose $||L||_{\infty}$ to simplify future calculations, see Appendix \ref{apx:ss} and \ref{apx:ss_c}. We note briefly that our GitHub page hosts the code necessary to re-implement the numerical portions of our study with $||L||_{2}$, which leads to very similar results and identical conclusions to what we find using $||L||_{\infty}$.}}  --  $||\mathbf{x}||_{\infty} = \max\{|x_1|, |x_2|, ... , |x_N|\}$ where $\mathbf{x}$ is the difference between two points in $\mathcal{M}$. We define a family of $h$ hash functions as $\mathcal{F}$ where for any two points $q_i$ and $q_j$ in $\mathcal{M}$ and any hash function $h$ chosen uniformly at random from $\mathcal{F}$ the following conditions hold:
\begin{align}
\label{eqn:lhs}
 \mathrm{if} \, \, \, d(q_i, \, q_j) \leq R&: \,  Pr[h(q_i) = h(q_j)] \geq p_1 \\
 \mathrm{if} \, \, \, d(q_i, \, q_j) > cR&: \, Pr[h(q_i) = h(q_j)] \leq p_2 \nonumber 
\end{align}
where threshold $R >0$, approximation factor $c>1$, and $Pr[]$ computes probabilities $p_1$ and $p_2$ with $0 \leq p_1, p_2 \leq 1$. \revs{In this work, we will store hash values $h(q_i)$ with the \texttt{numpy.float64} data type, thus $Pr[h(q_i) = h(q_j)] \approx 0$. In physical implementations of these systems, the precision of $h(q_i)$ will depend on the choice of sensor. Therefore, in establishing our mechanical analogue to a locality sensitive hashing algorithm, we re-write eqn. \ref{eqn:lhs} in terms of a positive value $S$ as:
\begin{align}
\label{eqn:lhs_v2}
 \mathrm{if} \, \, \, d(q_i, \, q_j) \leq R&: \,  Pr[d\big(h(q_i), \, h(q_j)\big) < S] \geq p_1 \\
 \mathrm{if} \, \, \, d(q_i, \, q_j) > cR&: \, Pr[d\big(h(q_i), \, h(q_j)\big) < S] \leq p_2 \nonumber 
\end{align}
which is elaborated on in Appendix \ref{apx:ss}.} With this definition, a family $\mathcal{F}$ is referred to as a ``Locality Sensitive Hash'' family, or alternatively as ($R$, $cR$, $p_1$, $p_2$)-\textit{sensitive} if $p_1 > p_2$. In simple and functional terms, illustrated schematically in Fig. \ref{fig:fig1}b, a family of hash functions will exhibit LSH behavior if the probability of a hash collision is higher for points that are closer together in the input space.

\subsection{Introduction to mechanical systems as locality sensitive hash functions}
\label{sec:methods_ss}

In this paper, we will explore the idea of using mechanical behavior to perform locality sensitive hashing where $\mathcal{F}$ will define a class of mechanical systems. In Fig. \ref{fig:fig1}c, we schematically illustrate our approach to defining this problem. Specifically, we will consider a vertical distributed load $w(x)$ applied on the surface of a mechanical system. \revs{The mechanical system will be} drawn from a family $\mathcal{F}$, where the mechanical behavior of the system \revs{will lead} to multiple force sensor readouts at discrete locations, treated as hash values. 

For this setup, we define the input continuous distributed load $w(x)$ as an evenly spaced $N \times 1$ dimensional vector (i.e., $w(x)$ is a continuous interpolation of $N$ points) and the process of ``hashing'' entails converting this $N$ dimensional vector into $ns$ (number of sensors) force sensor readouts. In future work the distributed load could be conceptualized as either multi-dimensional, i.e., $w(x,y)$, or displacement driven, and the sensor readouts could capture alternative forms of behavior (e.g., strain). 

\begin{figure}[h!]
	\centering
	\includegraphics[width=0.3\textwidth]{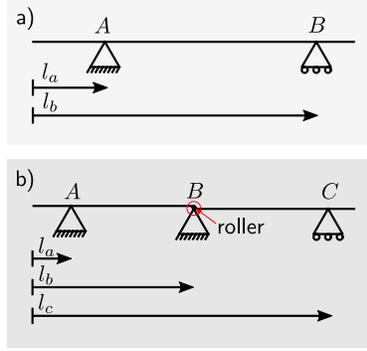}
	\caption{a) Schematic illustration of $\mathcal{F_{\mathrm{ss}}}$, \revs{a family of simply supported beams with two supports ($A$ and $B$)}; b) Schematic illustration of $\mathcal{F_{\mathrm{ss-c3}}}$, \revs{a family of simply supported composite beams with three supports ($A$, $B$, and $C$)}.}
	\label{fig:fig2}
\end{figure}

Here we will define our first family of mechanical hash functions $\mathcal{F}_{ss}$ as a family of simply supported beams, illustrated in Fig. \ref{fig:fig2}a, where supports $A$ and $B$ are randomly placed at positions $l_a$ and $l_b$ such that they are separated by a minimum distance $mL$ where $0<m<1$ and $L$ is the length of the beam. Here, $w(x)$ will be the input distributed load, and force at each of the $ns$ sensors is the hash value output. \revs{Following eqn. \ref{eqn:lhs_v2},} we will treat two hash values as a collision if the readouts at all sensors are within a tolerance $S$. We can then assess if the conditions defined in eqn. \ref{eqn:lhs}-\revs{\ref{eqn:lhs_v2}} hold for $\mathcal{F}_{ss}$. In Appendix \ref{apx:ss}, we expand on this in detail and explicitly define $R$, $cR$, $p_1$, and $p_2$ for $\mathcal{F}_{ss}$. However, for $\mathcal{F}_{ss}$, it only requires a simple thought experiment to demonstrate that $\mathcal{F}_{ss}$ is \textit{not} ($R$, $cR$, $p_1$, $p_2$)-\textit{sensitive}. 

In brief, we can demonstrate that $\mathcal{F}_{ss}$ is not ($R$, $cR$, $p_1$, $p_2$)-\textit{sensitive} by showing that we can choose two points in our inputs space ($w_1$, $w_2$), that are arbitrarily far apart ($d(w_1, w_2) > cR$ for an arbitrarily large $cR$) yet still lead to a hash collision for all possible mechanical hash functions in $\mathcal{F}_{ss}$ ($p_2 = 1$). In the context of mechanics, because all distributed loads with the same resultant force and centroid lead to the same reaction forces, simply supported beams do not meet the definition for locality sensitive mechanical hash functions.

However, if we consider even a slightly more complicated family of mechanical systems, simply supported composite beams with $3$ supports, referred to as $\mathcal{F}_{ss-c3}$ and illustrated in Fig. \ref{fig:fig2}b, the situation changes. 
For $\mathcal{F}_{ss-c3}$, we consider supports $A$, $B$, and $C$ that are randomly placed at positions $l_a$, $l_b$, and $l_c$ such that they are each separated by a minimum distance $mL$ where $0<m<0.5$ and segments $AB$ and $BC$ and connected through a roller support.
Because $l_a$, $l_b$, and $l_c$ change for each hash function in $\mathcal{F}_{ss-c3}$, two far apart distributed loads will collide with $p_2 < 1$, thus if we define $R$ as small enough such that the readout at each sensor will be within tolerance $S$ and thus $p_1 = 1$, we can show that $\mathcal{F}_{ss-c3}$ is ($R$, $cR$, $p_1$, $p_2$)-\textit{sensitive}. 
An explicit computation of $R$, $c$, $p_1$, and $p_2$ for $\mathcal{F}_{ss-c3}$ is expanded on in Appendix \ref{apx:ss_c}.
Overall, this simple demonstration is a proof of concept for mechanical systems as locality sensitive hash functions. 
Though this framework is straightforward, it is important to define as this work will lay the foundation for addressing more complex problems where we can consider broader definitions of mechanical hash functions, inputs, outputs, and system structure.
In Section \ref{sec:res_disc}, we will examine the functional performance of these simple beams alongside more complicated mechanical systems that may lead to more desirable functional LSH behavior. \revs{In addition, the straightforward LSH approach defined here will serve as a baseline for novel strategies to optimize mechanical systems for downstream signal processing.}

\subsection{Example problem definition}
\label{sec:example_problem}

Beyond satisfying the criteria defined in eqn. \ref{eqn:lhs}, we are interested in assessing the functional utility of mechanical systems for performing locality sensitive hashing. 
To this end, we now define an example problem consisting of example loading, defined in Section \ref{sec:problem_loads}, example
mechanical systems, defined in Section \ref{sec:problem_sys}, and evaluation metrics, defined in Section \ref{sec:problem_metrics}. In brief, we will investigate how different mechanical systems transform mechanical signals in the context of LSH.

\begin{figure}[p]
	\centering
	\includegraphics[width=.975\textwidth]{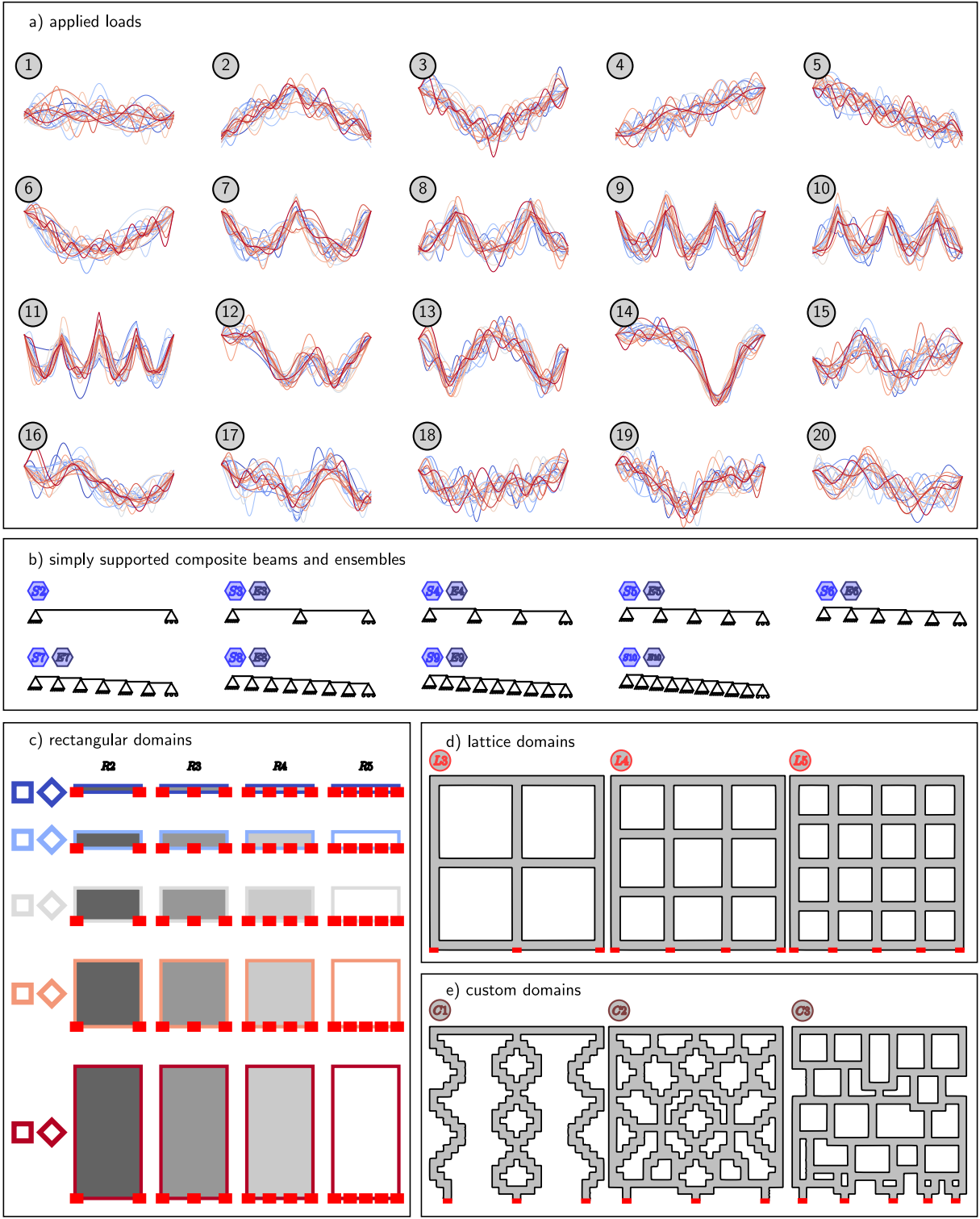}
	\caption{a) Illustration of the $20$ classes of applied loads; b) Schematic illustration of the simply supported and ensemble beam mechanical systems (note an example ensemble is explicitly illustrated in Appendix Fig. \ref{fig:fig3_apx}); c) Schematic illustration of the rectangular domains with different depths ($depth = [1.0, 2.5, 5.0, 10.0, 20.0]$) and number of sensors ($ns = [2, 3, 4, 5]$). For all rectangular domains, we simulate the bottom fixed only at sensors ($\square$) and the whole bottom fixed ({\Large $\diamond$}); d) Schematic illustration of the lattice domains ($L3$, $L4$, $L5$); e) Schematic illustration of the custom domains ($C1$, $C2$, $C3$). \revs{Note that all lattice and custom domains have $depth=10.0$.} In b-e, the markers next to each mechanical system match the markers used in Fig. \ref{fig:fig4}-\ref{fig:fig5}.}
	\label{fig:fig3}
\end{figure}

\subsubsection{Example loads}
\label{sec:problem_loads}

In Fig. \ref{fig:fig3}a, we show our $200$ randomly generated applied loads across $20$ different \revs{categorical classes}. In brief, we consider: constant loads, linear loads, piecewise linear loads, absolute value sinusoidal loads, and kernel density estimate loads based on randomly generated point densities. Note that, as illustrated in Fig. \ref{fig:fig1}c, we apply all loads pointing downwards. Building on the framework laid out in Section \ref{sec:methods_ss}, these loads are chosen to: (1) pose a risk of unwanted hash collisions across categories (i.e., identical centroids and resultant forces), while (2) being qualitatively different. For each of the $20$ \revs{categorical classes}, we generate $20$ examples with different selections of \textit{correlated} Perlin noise \citep{perlin1985image,perlinPython}. Specifically, we add Perlin noise with a randomly selected seed and octave (random integer with range $[2,10]$) -- this is the source of the variation across each category in the curves illustrated in Fig. \ref{fig:fig3}a. Additional details for describing all categories of load are given in Appendix \ref{apx:alc}, and the link to the code for re-generating these loads including selected random seeds is given in Section \ref{sec:additional_info}. 

\subsubsection{Example mechanical systems}
\label{sec:problem_sys}

In Fig. \ref{fig:fig3}b-e, we show the mechanical systems investigated in this study. For all examples, we set the top surface length $L=10$ (length units). In brief, we consider the following classes of mechanical systems:
\begin{itemize}
\item Simply supported and simply supported composite beams defined in Section \ref{sec:methods_ss}. We will consider composite beams with up to $10$ supports, and we will report the performance of both single beam instances and hard voting based ensemble behavior for a total of $17$ different scenarios (see Fig. \ref{fig:fig3}b). For the beam ensembles, each of the $100$ composite beams will have different randomly generated support locations. The final performance of each ensemble will then be represented as a single value that combines information from all $100$ randomly generated composite beams. An illustration of a representative beam ensemble is included in Appendix \ref{apx:ens}. 
\item Homogeneous rectangular domains with variable depth, number of sensors, and fixity. We will consider rectangular domains with $2$, $3$, $4$, and $5$ force sensors, depths $1.0$, $2.5$, $5.0$, $10.0$, and $20.0$, and both with and without bottom fixity. \revs{For reference, a rectangular domain with $depth=1.0$ will have dimension $1.0 \times 10.0$, and a rectangular domain with $depth=10.0$ will be a square.} \revs{The combination of $4$ sensor options, $5$ depths, and $2$ bottom fixities leads to} leads to a total of \revs{$4 \times 5 \times 2 = 40$} different \revs{rectangular} domains (see Fig. \ref{fig:fig3}c \revs{where all potential sensor placements and depths are illustrated}).
\item Lattice domains with $3$, $4$, and $5$ sensors and corresponding $2 \times 2$, $3 \times 3$, and $4 \times 4$ window grids (see Fig. \ref{fig:fig3}d). \revs{All lattice domains have $depth=10.0$.}
\item Custom domains with three different geometries and variable sensor numbers $ns$ (see Fig. \ref{fig:fig3}e). \revs{All custom domains have $depth=10.0$.}
\end{itemize}
For each mechanical system, we run $400$ simulations corresponding to each of the applied loads illustrated in Fig. \ref{fig:fig3}a and report the $y$ direction force at each sensor location. In the context of LSH, the applied loads are the input, and these forces are the hash function output. In all cases, sensor locations are fixed in both the $x$ and $y$ direction, but only the $y$ direction force is used in subsequent analysis.
Simply supported beams are simulated in a Python \citep{2020NumPy-Array} script that we link to in Section \ref{sec:additional_info}.
All finite element simulations are conducted using open source finite element software FEniCS \citep{logg2012automated,alnaes2015fenics} and built in FEniCS mesh generation software \texttt{mshr}. To avoid numerical artifacts, we simulate all domains as Neo-Hookean materials with $\nu=0.3$, and a fine mesh (\texttt{mshr} mesh parameter set to $200$) of quadratic triangular elements.
The link to the code for re-generating all simulation results is given in Section \ref{sec:additional_info}. In addition, our code contains a tutorial for designing and simulating user defined architected domains to make it straightforward to expand on the initial results presented in this paper.

\subsubsection{Evaluation metrics}
\label{sec:problem_metrics}

Based on the suite of $400$ loads defined in Section \ref{sec:problem_loads} and the $63$ mechanical systems defined in Section \ref{sec:problem_sys} (see Fig. \ref{fig:fig3}), we will have over $25,000$ simulation results to analyze. To draw conclusions from these results, we will examine the relationship between three different quantities of interest: \revs{(1)} the probability of a hash collision with respect to the $L^{\infty}$ distance between loads, \revs{(2)} the Spearman's Rank correlation coefficient $\rho$ between the $L^{\infty}$ distance between loads and the $L^{\infty}$ distance between hash values \citep{myers2013research,2020SciPy-NMeth}, and (3) the classification accuracy based on a single nearest neighbor (note that there are $20$ categories of loads, with $20$ examples in each category, illustrated in Fig. \ref{fig:fig3}a) \citep{pedregosa2011scikit}.

\begin{figure}[h]
	\centering
        \includegraphics[width=\textwidth]{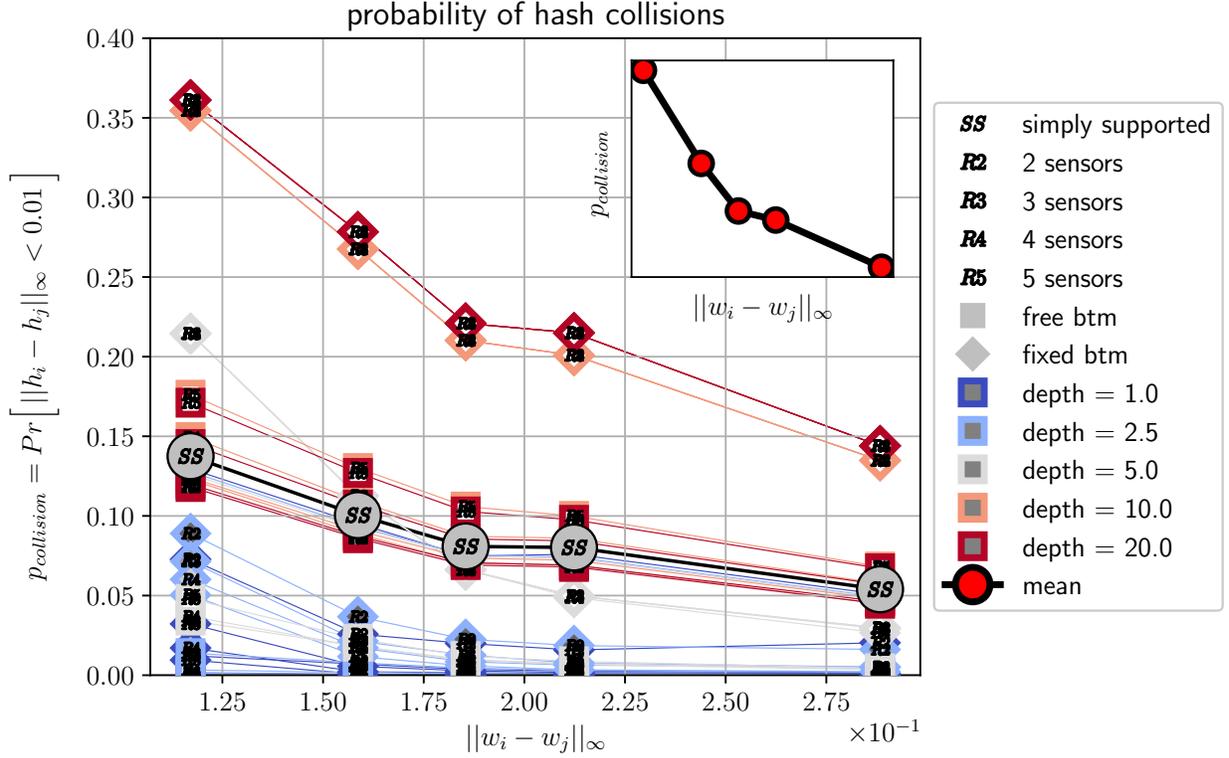}
	\caption{\revs{Plot of $p_{\mathrm{collision}}$ of hash values (defined as the probability that $|| h_i - h_j||_{\infty} < 0.01$) vs. distance between input loads $|| w_i - w_j||_{\infty}$ for all rectangular domains. For context, we also include the curve for a simply supported beam with two supports, and the curve for the mean of all rectangular samples (see inset, where $p_{\mathrm{collision}}=[0.14, 0.099, 0.075, 0.070, 0.047]$).}}
	\label{fig:fig4}
\end{figure}

To \revs{begin}, we will report the probability of hash collision $p_{collision}$ as function of the $L^{\infty}$ distance between input loads \revs{and hash values} for all $40$ rectangular domains \revs{and compare to a baseline simply supported beam from $\mathcal{F}_{ss}$}.
To \revs{approximate} $p_{collision}$ \revs{vs. $L^{\infty}$ distance for our selection of example loads}, we divide all load pairs into five equally sized bins based on the distance between input loads. Then, we compute $p_{collision}$ as the fraction of hash values in each bin with $L^{\infty}<0.01$ (input loads are normalized to sum up to $1$). \nr{In Fig. \ref{fig:fig4}, we plot $p_{collision}$ with respect to the average input distance value associated with each equally sized bin.} Thus, for each device we will have a curve that shows binned $p_{collision}$ vs. $L^{\infty}$ distance between loads. In Section \ref{sec:res_disc}, we visualize this behavior in Fig. \ref{fig:fig4} as one \revs{(limited)} approach to observing LSH behavior. 

\revs{Beyond $p_{collision}$ vs. $L^{\infty}$ distance,} we will also compute Spearman's $\rho$ designed to capture the rank correlation between the $L^{\infty}$ distance between loads and the $L^{\infty}$ distance between hash values. For all $79,800$ load \textit{pairs} ($400 \times 399/2 = 79,800$ \textit{pairs}) we rank order the distances of both \nr{the loads and the hash values} and then compute $\rho$ as:
\begin{equation}
\ \rho = 1 - \frac{6 \sum\limits_{i=1}^{n} r_i^2}{n \, (n^2 -1)}
\end{equation}
where \revs{$r_i$} is the difference between the \nr{ranks of the load and hash values distances and $n$ is the number of load pairs}. For perfect monotonic correlation $\rho = 1$ and for no correlation $\rho = 0$. We perform this operation with the Python function \texttt{scipy.stats.spearmanr()} \citep{2020SciPy-NMeth}. \revs{Though visualizing $p_{collision}$ vs. $L^{\infty}$ is a more intuitive match to the LSH definitions in eqn. \ref{eqn:lhs}-\ref{eqn:lhs_v2}, the quantitative comparison of preserving input relationships is perhaps more interpretable via Spearman's $\rho$.} In Section \ref{sec:res_disc}, we visualize Spearman's $\rho$ in each device both with respect to \revs{classification accuracy} (Fig. \ref{fig:fig5}), and \revs{mechanical system properties} (Fig. \ref{fig:fig6}, see also Appendix Fig. \ref{fig:fig12}). 

The third evaluation quantity is \revs{chosen as a set up for future applications in ``learning to hash.'' 
\nr{In future ``learning to hash'' applications, the hashing efficacy of the mechanical system will be measured via the performance on a functional task. Here we choose classification accuracy as an example functional task. In brief, each of the $400$ loads belongs to one of $20$ applied load classes ($20$ classes, $20$ loads per class). This is illustrated in Fig. \ref{fig:fig3}a. 
Here, we compute} classification accuracy using a simple nearest neighbor algorithm \nr{where we predict the class of a given load based on its hash value.} For each of the $400$ loads, we implement a new k-nearest neighbor classifier based on the $399$ other loads and then see what class is predicted for the held out load with $k=1$ \citep{james2013introduction}.
Classification accuracy is then defined based on the ability of this algorithm to predict the class of the held out load:}
\begin{equation}
    \mathrm{accuracy} = \frac{\mathrm{correct \, predictions}}{\mathrm{total \, predictions}}
    \label{eqn:acc}
\end{equation}
where the ``correct'' prediction is which of the $20$ applied load classes (see Fig. \ref{fig:fig3}a) the input load belongs to. We implement this algorithm with scikit-learn \citep{pedregosa2011scikit} and report the average prediction accuracy across all $400$ individual hold out cases. Because there are $20$ identically sized \nr{labeled} classes \nr{according to the problem definition}, the baseline prediction accuracy that represents random guessing is $0.05$. In Section \ref{sec:res_disc}, classification accuracy is reported in Fig. \ref{fig:fig5}.
As a brief note, the link to the code for computing all of these quantities of interest is given in Section \ref{sec:additional_info}.

\section{Results and Discussion}
\label{sec:res_disc}

In this Section, we will summarize the results of the investigation detailed in Section \ref{sec:example_problem}. On one hand, the results of this study are largely intuitive -- applied loads influence mechanical response, and different applied loads lead to different mechanical response. On the other hand, this initial investigation is an important step because it lays the groundwork for significant future investigation in \revs{designing mechanical systems that outperform the baseline proof of concept results shown here}. Multiple future directions are explicitly stated in Section \ref{sec:conclusion}.

\subsection{\revs{Probability of hash collision decreases with increasing distance between input loads}}
\label{sec:res_corr}

The first major result of this investigation is shown in Fig. \ref{fig:fig4} where we visualize the probability of a hash collision $p_{\mathrm{collision}}$ vs. the $L^{\infty}$ distance between the normalized input loads (sampled with $N=1000$) for all rectangular domains. Following Section \ref{sec:problem_sys} and Section \ref{sec:example_problem}, we define a collision as the circumstance where every component of the hash value is within distance $S=0.01$ (with all input loads normalized to have the same total resultant force). For example, $[0.5, 0.25, 0.25]$ and $[0.495, 0.2525, 0.2525]$ would ``collide.'' 
The critical outcome shown in this plot is that $p_{\mathrm{collision}}$ decreases as the $L^{\infty}$ distance increases, which is desirable behavior for \revs{locality sensitive hashing and hashing for similarity search in general}. 
\revs{We note briefly that the inset plot of Fig. \ref{fig:fig4} shows the mean $p_{\mathrm{collision}}$ curve for all rectangular domains where this decrease is readily visible}.
\revs{However, in Fig. \ref{fig:fig4}, we also plot a baseline $p_{\mathrm{collision}}$ curve for a simply supported beam with two supports which shows that observing a decrease in $p_{\mathrm{collision}}$ vs. the $L^{\infty}$ input distance for this selection of input loads (see Fig. \ref{fig:fig3}a) is not sufficient to claim LSH behavior. Therefore, we turn to the other metrics defined in Section \ref{sec:problem_metrics}, Spearman's $\rho$ and classification accuracy, to add needed context to our investigation.}

\begin{figure}[h]
	\centering
	\includegraphics[width=\textwidth]{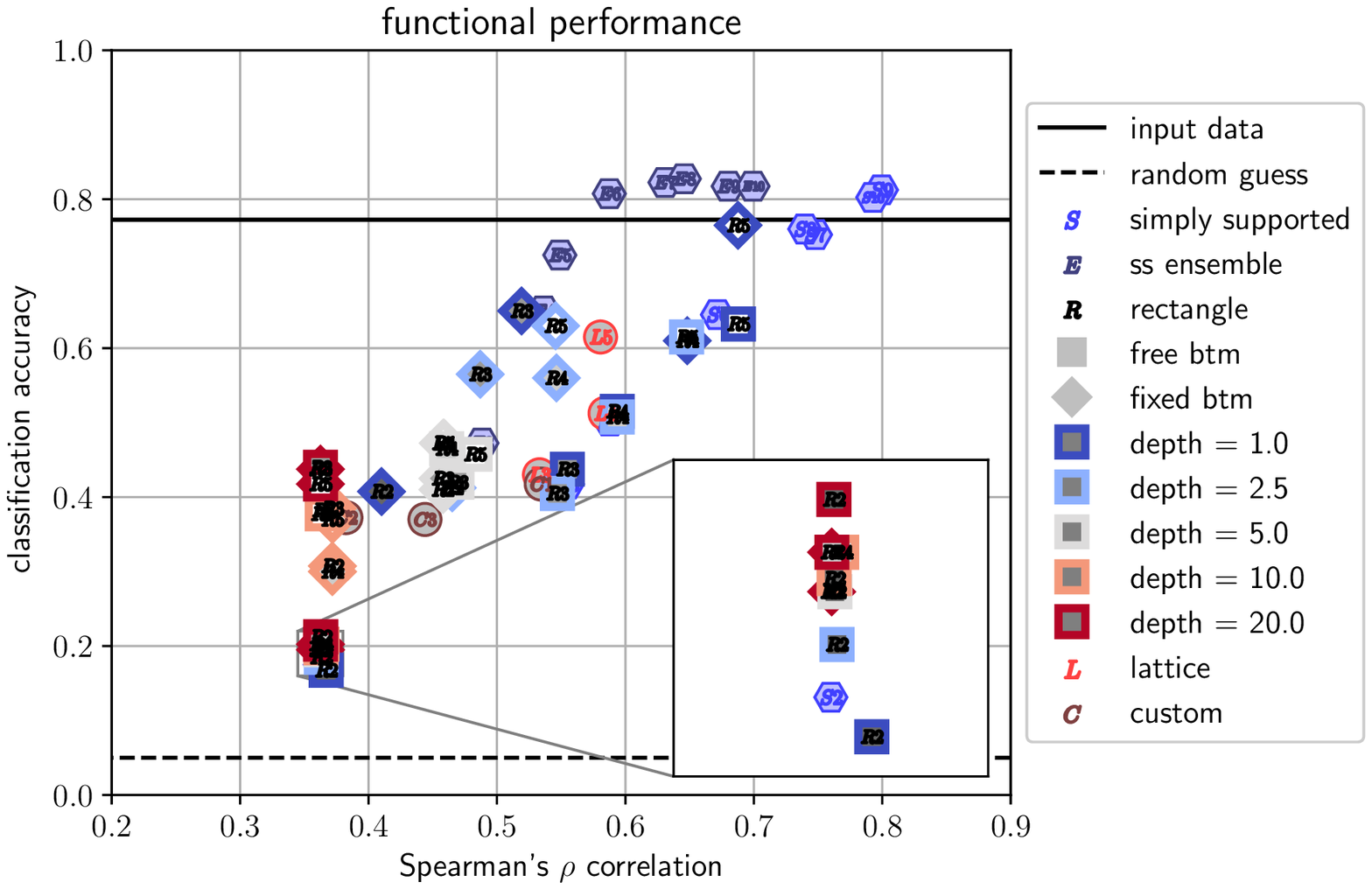}
	\caption{\revs{Plot of classification accuracy vs. Spearman's $\rho$ for all mechanical systems explored in this study. Note the inset plot which contains $S2$ (simply supported beam with two supports) and similarly performing mechanical systems. Fig. \ref{fig:fig12} also presents Spearman's $\rho$ and classification accuracy plotted with respect to number of sensors and domain depth.}}
	\label{fig:fig5}
\end{figure}

\subsection{\revs{Spearman's $\rho$ and classification accuracy vary across mechanical systems}}
\label{sec:res_acc}

\revs{In Fig. \ref{fig:fig5}, we explore two key components of the \textit{functional} behavior of our mechanical hashing systems, Spearman's $\rho$ and classification accuracy, and visualize their relationship.
From a \textit{functional} perspective, Spearman's $\rho$ captures a picture of the potentially non-linear distance preservation between the input loads and the hash values. And, from a \textit{functional} perspective, assessing classification accuracy will set up a toy problem and baseline functional performance for future work in \textit{optimizing} mechanical domains to perform hashing functions.  Specifically, we anticipate future work in designing application specific mechanical systems that target specific functionality (e.g., high classification accuracy) where the results shown here will serve as baselines at these tasks.  
}

Overall, we observe that accuracy ranges from $\approx 0.17$, the $accuracy$ of a simply supported beam with two supports, to $\approx 0.8$, the accuracy for simply supported ensembles with $>5$ sensors (note that force sensors are all located at the supports). For reference, $accuracy=0.05$ corresponds to random guessing, and $accuracy=0.77$ corresponds to the \nr{prediction accuracy that is obtained by performing classification with the input loads directly rather than with the hash values}. And, overall, Spearman's $\rho$ and classification accuracy appear to be correlated, which is consistent with LSH behavior for scenarios where the hash function is not specifically learned to perform a non-linear transformation on the distribution of inputs.  In addition, it is worth mentioning that the inset plot in Fig. \ref{fig:fig5} contains multiple ``poorly performing'' domains with both low Spearman's $\rho$ and low classification accuracy. Notably, the domains in this region are rectangular domains that are deep and/or have only two sensors. And, as demonstrated by the inset plot, these domains all perform similarly to the simply supported beam with two supports.

\revs{For each of the mechanical systems introduced in Section \ref{sec:problem_sys}, key observations are as follows:
\begin{itemize}
    \item For the simply supported composite beams and beam ensembles, increasing $ns$ leads to both higher Spearman's $\rho$ and higher $accuracy$. And, the hard voting based ensemble predictions tend to outperform the single mechanical system results. Notably, the transition between $ns=2$ and $ns=3$ leads to the largest incremental improvement in performance for the simply supported beams ($accuracy=0.175$ to $accuracy=0.42$), which is also consistent with our introduction to the concept of LSH in Section \ref{sec:methods_ss}.
    \item For the rectangular domains, $accuracy$ and Spearman's $\rho$ vary widely. As stated previously, some designs perform poorly -- similar to the simply supported beam with $ns = 2$ -- whereas other designs exceed the simply supported composite beam performance for the equivalent number of sensors. In general, increasing $ns$ and decreasing $depth$ correspond to increases in $\rho$ and improvements in $accuracy$. From a mechanics perspective, this is a logical result as increasing $ns$ will provide more information about mechanical behavior while increasing $depth$ will lead to diminishing differentiation between applied tractions with the same resultant force and centroid following Saint-Venant’s principle \citep{bower2009applied}. 
    \item In comparison to the rectangular domains, the lattice domains led to consistent performance improvements ($accuracy_{L3}=0.43$, $accuracy_{L4}=0.51$, $accuracy_{L5}=0.61$).
    \item In comparison to the rectangular domains and the lattice domains, the custom domains selected offered little performance improvement ($accuracy_{C1} = 0.42$, $accuracy_{C2} = 0.37$, $accuracy_{C3} = 0.37$). 
\end{itemize}
}

\revs{Further visualizations are shown in Fig. \ref{fig:fig12} where Spearman's $\rho$ and classification accuracy are plotted with respect to the number of sensors and domain depth. And, in Table \ref{tab:tab1}, we also list Spearman's $\rho$ and classification accuracy for each domain directly.} Finally, it is worth re-emphasizing that even when classification is performed on the input signals directly, we only achieve $accuracy=0.77$. This is because we defined an example problem with loads that can be difficult to disaggregate in the presence of noise. In Appendix \ref{sec:res_apx} Fig. \ref{fig:fig6}, we provide the confusion matrix for the unaltered input signals to highlight which loads are leading to overlapping predictions. This quantitative outcome is consistent with the qualitative comparison that can be made by examining the plots in Fig. \ref{fig:fig3}a.

\begin{figure}[h]
	\centering
	\includegraphics[width=.8\textwidth,trim={0mm 0mm 0mm 0mm},clip]{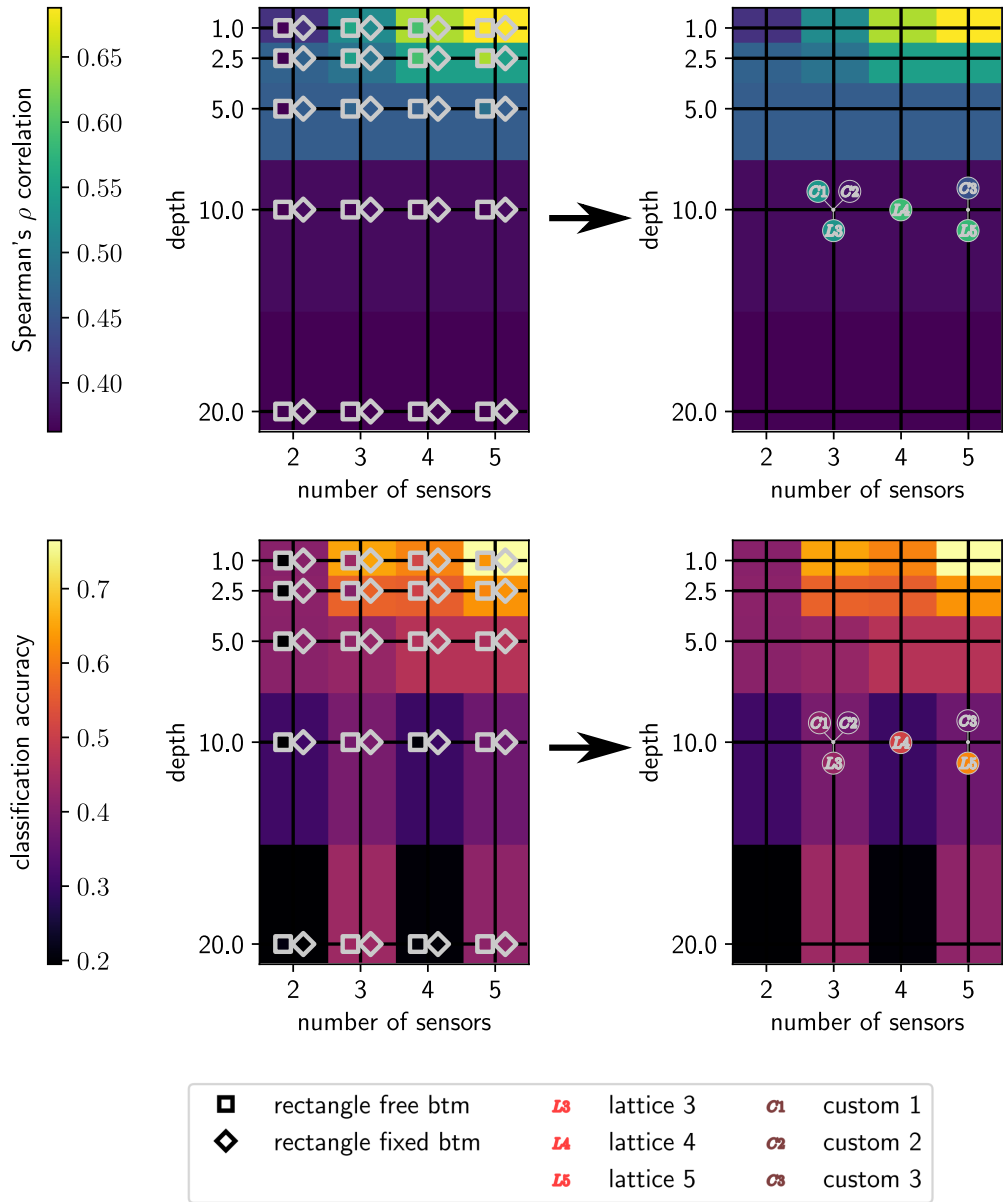}
	\caption{\revs{Visualization of Spearman's $\rho$ (upper) and classification accuracy (lower) with respect to domain depth and number of sensors. In the left column, results from individual rectangular domains are indicated by the {\footnotesize $\square$} and {\large $\diamond$} markers, and the background shading is based on the rectangular domain with a fixed bottom. In the right column, results from the lattice and custom domains are superimposed on the same background shading, thus comparing the lattice and custom domains to the rectangular domain baseline. Note that the fill color of all markers is dictated by Spearman's $\rho$ (upper) and classification accuracy (lower). Figure \ref{fig:fig12} also presents Spearman's $\rho$ and classification accuracy plotted with respect to number of sensors and domain depth. }}
	\label{fig:fig6}
\end{figure}

\subsection{\revs{Architected domains change, and can be used to enhance, task specific LSH performance}}
\label{sec:arch}

\revs{In Fig. \ref{fig:fig6}, we re-organize the rectangular, lattice, and custom domain data shown in Fig. \ref{fig:fig5} to better visualize the influence of architected domains on Spearman's $\rho$ and classification accuracy with respect to both domain depth and number of sensors. Critically, this figure demonstrates that even for a fixed number of sensors, there is a large variation in domain performance. And, by comparing Spearman's $\rho$ and classification accuracy of the lattice domains to the rectangular domains, we can see that architecting these domains can help overcome the drop off in performance with respect to domain depth. Finally, the spread in performance between the lattice and custom domains also indicates that there is potential richness to this problem where selecting domains that will perform well is non-trivial. These observations taken together point towards a strong future opportunity to engineer systems that, for a given suite of possible loads and allowable number of sensors, are specifically designed to maximize Spearman's $\rho$, classification accuracy, and/or achieve an alternative type of engineered relationship between mechanical inputs and sensor readouts. Though the notion that architected domains can alter force transmission is a straightforward result, the framework that we have established here will directly enable future work in the design and optimization of architected domains to perform desirable application specific signal transformations.}

\section{Conclusion}
\label{sec:conclusion}

\revs{In this paper, we began with a brief introduction to the concept of hashing and hashing for similarity search. Then, we define locality sensitive hashing and lay the foundation for considering mechanical systems as locality sensitive hash functions. From both our analytical and computational investigations, we find that mechanical systems can exhibit the properties required for locality sensitive hashing, and we find that by tuning mechanical inputs (e.g., boundary conditions, domain architecture) we can change the \nr{functional} efficacy of mechanical systems for this task. Based on our observations, and the very general scope of ``mechanical systems,'' we anticipate that there will be a significantly broader potential range of behavior than what we captured in the systems selected for this study. Overall, the main contributions of this work are to: (1) introduce the concept of locality sensitive hashing via mechanical behavior, \nr{(2) define a numerical approach to readily assessing metrics that indicate functional locality sensitive hashing behavior}, and (3) establish a baseline performance for future comparison where mechanical systems are optimized to perform hashing for similarity search related tasks.}

\revs{Following this thread,} is worth highlighting that we view this investigation as a starting point for significant further study of hashing performed by mechanical systems.
Looking forward, we anticipate \nr{four} key endeavors that will build on this work. \nr{First, future work is required to demonstrate that mechanical systems besides simply supported and simply supported composite beams do or do not meet the \textit{formal} definition of locality sensitive hash functions. We anticipate that this work will be conducted by beginning with our straightforward to compute \textit{proxies} for locality sensitive behavior, and then showing formally that a given mechanical system is able to meet the definition laid out in eqn. \ref{eqn:lhs_v2} by anticipating extreme load pairs following the procedure in Appendix \ref{apx:ss}.} Second, an important next step is the physical realization of mechanical systems for locality sensitive hashing. Constraints imposed by the need for ready constructability and current sensing capabilities will pose a challenge \citep{tapia2020makesense}, and it will be important to determine if our findings remain consistent in the equivalent experimentally realized systems.
Third, it will be interesting to explore the efficacy of a \textit{learning to hash} approach in mechanical systems where the input distribution is known and the hash function (i.e., the mechanical system) is designed specifically to perform a desired task \citep{wang2017survey,wang2015learning}. The framework and metrics defined here will allow us to construct an optimization problem where both system mechanical behavior and sensor placement can be jointly tailored to serving a specific function. In future learning to hash applications, the mechanical systems explored in this work can serve as a baseline for comparison, where optimized systems should lead to better performance of a desired task. 
Because this is a challenging optimization problem, we anticipate that there will be a need to implement efficient modeling and optimization strategies \citep{prachaseree2022learning, mohammadzadeh2022predicting,bessa2019bayesian, senhora2022machine}.
Finally, we anticipate that that the structural form of architected materials that are highly effective at hashing for similarity search may exist in nature, and identifying relevant motifs may help us better understand force transmission in biological cells and tissue. 
Though this initial study is quite straightforward, we anticipate that it will directly enable a highly novel approach to physical computing. 

\section{Additional Information}
\label{sec:additional_info}

The data and code to reproduce and build on the results in this paper are provided on our GitHub page (\url{https://github.com/elejeune11/mechHS}).

\section{Acknowledgements}

This work was made possible with funding through the Boston University David R. Dalton Career Development Professorship, the Hariri Institute Junior Faculty Fellowship, the Haythornthwaite Foundation Research Initiation Grant, and the National Science Foundation Grant CMMI-2127864. This support is gratefully acknowledge.

\appendix

\section{Simply Supported Composite Beams as Locality Sensitive Hash Functions}
\label{apx:mech_HS}

Here we provide supporting information for the work presented in Section \ref{sec:meth_lsh}. As stated in the main text, we are exploring the problem schematically illustrated in Fig. \ref{fig:fig1}c. Following the definition in eqn. \ref{eqn:lhs}, our goal is to determine if a given family of mechanical hash functions $\mathcal{F}$ is ($R$, $cR$, $p_1$, $p_2$)-\textit{sensitive}. 

\subsection{Simply Supported Beams}
\label{apx:ss}

\begin{figure}[h!]
	\centering
	\includegraphics[width=.7\textwidth]{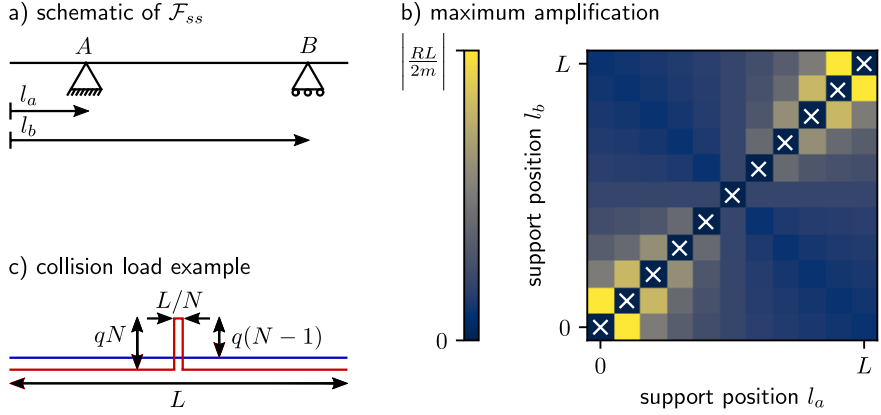}
	\caption{\revs{a) Schematic illustration of $\mathcal{F_{\mathrm{ss}}}$; b) Visualization of the amplification of the $\pm R \, \mathbf{1}$ term defined in eqn. \ref{eqn:w2_coll} for different support positions $l_a$ and $l_b$ (here $m=0.1$, the $\times$ markers indicate choices where $|l_a - l_b| < mL$, and we discretize $11$ evenly spaced potential support positions $[0, L]$); c) Schematic illustration of an example of two loads that will always collide for $\mathcal{F_{\mathrm{ss}}}$.}}
	\label{fig:fig2_apx}
\end{figure}

As our first exploration, we define a family of mechanical hash functions $\mathcal{F}_{ss}$ as a family of simply supported beam reaction forces. This family, illustrated in Fig. \ref{fig:fig2_apx}\revs{a}, is defined by the randomly generated placement of reaction supports, $A$ and $B$ placed at $l_a$ and $l_b$ respectively. These supports can have any location as long as they are separated by distance $mL$, where $0<m<1$ and $L$ is the length of the beam. To begin, we will first establish $R$, the threshold distance, and $p_1$, the probability of a collision for two loads within the threshold distance.  For simplicity, we choose to define $R$ for $p_1=1$. To do this, we need to define the threshold distance between two loads $w(x)_1$ and $w(x)_2$ that will always hash to the same value, defined as $h[w_1(x)] = h[w_2(x)]$ and alternatively written as $h_1=h_2$. To explicitly define distance $R$, we need to define the size of our hash buckets $S$. Because our outputs will come in the form of continuous numerical values, we will conceptualize our hash buckets as discrete bins that break up this continuous space into bins of size $S$. To mitigate the influence of the placement of the bin boundaries, we will simply consider any two numbers within distance $S$ as identical. Therefore, for a simply supported beam, two loads $w_1(x)$ and $w_2(x)$ will experience a hash collision when both reaction forces collide, defined as:
\begin{equation}
    |A_{y1} - A_{y2}| < S \qquad \mathrm{and} \qquad |B_{y1} - B_{y2}| < S \, .
\end{equation}
For $\mathcal{F}_{ss}$, we can define $w_1$ and $w_2$ in terms of $R$ as:
\begin{equation}
    w_2(x) = w_1(x) \pm R \, \mathbf{1}
    \label{eqn:w2_coll}
\end{equation}
where $\mathbf{1}$ is a vector with the same length $N$ as $w_1(x)$ and $w_2(x)$ such that $R$ is the $L^{\infty}$ norm of the distance between $w_1(x)$ and $w_2(x)$. 
For $\mathcal{F}_{ss}$, \revs{as illustrated in Fig. \ref{fig:fig2_apx}b}, the most extreme amplification of the $\pm R \, \mathbf{1}$ term will occur when $l_a=0$ and $l_b=mL$ (alternatively $l_a = L - mL$, $l_b$ = L). For $l_a=0$ and $l_b=mL$, we perform a simple equilibrium calculation to compute:
\begin{align}
     A_{y1} &= \int_0^L w_1(x) \, \mathrm{d}x - \frac{1}{mL} \int_0^L w_1(x) \, x \, \mathrm{d}x \nonumber \\
     B_{y1} &= \frac{1}{mL} \int_0^L w_1(x) \, x \, \mathrm{d}x  \\ 
     A_{y2} &= A_{y1} \pm \bigg(1 - \frac{1}{2m}\bigg) RL  \nonumber \\
     B_{y2} &= B_{y1} \pm \bigg(\frac{1}{2m}\bigg) RL \nonumber
\end{align}
which allows us to compute:
\begin{align}
    \bigg|A_{y1} - A_{y2}\bigg| &= \bigg|\bigg( 1 - \frac{1}{2m}\bigg) RL \bigg| \\
    \bigg|B_{y1} - B_{y2}\bigg| &= \bigg|\bigg(\frac{1}{2m}\bigg) RL\bigg| \nonumber
\end{align}
where:
\begin{align}
    \label{eqn:S_solve}
    S &= \max\bigg(\, \big|\big( 1 - \frac{1}{2m}\big) RL \big|, \, \big|\big(\frac{1}{2m}\big) RL\big|\, \bigg) \\ 
      &= \big|\big(\frac{1}{2m}\big) RL\big| \nonumber
\end{align}
for $0 < m < 1$ which allows us to compute:
\begin{equation}
    R=2mS/L \, .
\end{equation}
Therefore, for $d(p,q) < 2mS/L$, $p_1 = 1$. 

Following our identification of $R$ for $p_1 = 1$, we need to see if it is possible to specify $c$ such that $p_2 < p_1$. 
Here is where we encounter the fundamental limitation of $\mathcal{F}_{ss}$ as a ($R$, $cR$, $p_1$, $p_2$)-\textit{sensitive} hash function. 

Of course, for $\mathcal{F}_{ss}$, we can define two loads $w_1(x)$ and $w_2(x)$ that are arbitrarily far apart (i.e., no upper limit on $c$) that will always lead to a hash collision. Specifically, we just need to choose two different loads with the same resultant force and centroid. For example, illustrated in Fig. \ref{fig:fig2_apx}\revs{c}, we can define one load as:
\begin{equation}
    w_1(x) = q \mathbf{1}
\end{equation}
where $q$ is a constant and $\mathbf{1}$ is length $N$ vector of ones, and another load as a central spike, written as:
\begin{align}
    w_2(x) = 
    \begin{cases}
        0,& x < \frac{L(N-1)}{2N}\\
        qN,& \frac{L(N-1)}{2N} \leq x \leq \frac{L(N+1)}{2N}\\
        0,& \frac{L(N+1)}{2N} < x\\
    \end{cases}
\end{align}
and illustrated in Fig. \ref{fig:fig2_apx}\revs{c}. For this choice of $w_1(x)$ and $w_2(x)$, we can compute $|| w_1(x) - w_2(x)||_{\infty}$ as:
\begin{equation}
    || w_1(x) - w_2(x)||_{\infty} = q(N - 1)
\end{equation}
which can become arbitrarily large.\footnote{$|| w_1(x) - w_2(x)||_{\infty} = q(N - 1)$ if $N$ is odd, $|| w_1(x) - w_2(x)||_{\infty} = q(N/2 - 1)$ if $N$ is even.}
Because $p_2 = 1$ for any value of $c$, we can formally say that $\mathcal{F}_{ss}$ is \textit{not} ($R$, $cR$, $p_1$, $p_2$)-\textit{sensitive}. 

\subsection{Simply Supported Composite Beams with $>2$ Supports}
\label{apx:ss_c}

\begin{figure}[h!]
	\centering
	\includegraphics[width=.7\textwidth]{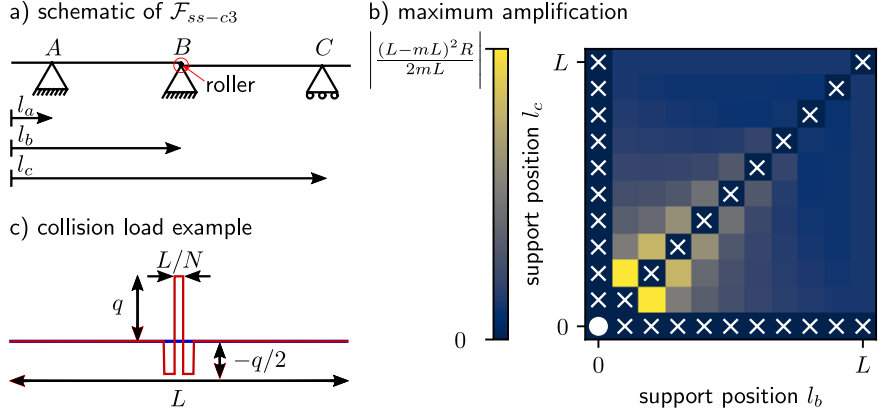}
	\caption{\revs{a) Schematic illustration of $\mathcal{F_{\mathrm{ss-c3}}}$; b) Visualization of the amplification of the $\pm R \, \mathbf{1}$ term defined in eqn. \ref{eqn:w2_coll} for different support positions $l_b$ and $l_c$ (here $m=0.1$, the $\bullet$ marker indicates the selection of $l_a=0$, the $\times$ markers indicate choices where $|l_a - l_b| < mL$ or $|l_a - l_c| < mL$ or $|l_b - l_c| < mL$, and we discretize $11$ evenly spaced potential support positions $[0, L]$); c) Schematic illustration of a pair of loads that we believe will have the highest probability of a hash collision for mechanical systems in $\mathcal{F_{\mathrm{ss-c3}}}$.}}
	\label{fig:fig3_apx}
\end{figure}

If we instead consider a slightly more complicated family of mechanical systems, simply supported composite beams with 3 supports, referred to as $\mathcal{F}_{ss-c3}$ and illustrated in Fig. \revs{\ref{fig:fig3_apx}a}, this picture changes. 
Here, we consider a composite beam where segments $AB$ and $BC$ are connected via a roller support with supports $A$, $B$, and $C$ located at $l_a$, $l_b$, and $l_c$ respectively. In this case, a hash collision between two loads $w_1(x)$ and $w_2(x)$ requires all three support reactions to collide, written as:
\begin{equation}
    |A_{y1} - A_{y2}| < S \qquad \mathrm{and} \qquad |B_{y1} - B_{y2}| < S \qquad |C_{y1} -C_{y2}| < S \, .
\end{equation}
We can follow the same logic to compute $R$ as the prior $\mathcal{F}_{ss}$ example. Specifically, for $\mathcal{F}_{ss-c3}$ we identify the the most extreme amplification of the $\pm R \mathbf{1}$ term introduced in eqn. \ref{eqn:w2_coll} for $l_a=0$, $l_b=mL$, and $l_c=2mL$, where $m$ controls the minimum allowable distance between supports \revs{(see Fig. \ref{fig:fig3_apx}b for a visualization of this amplification)}. Again, we perform a simple equilibrium calculation to compute the reaction supports $A_y$, $B_y$, $C_y$ for $w_1(x)$ and $w_2(x)$ as:
\begin{align}
    A_{y1} &= \frac{mL\int_0^{mL} w_1(x) \, \mathrm{d}x \,  - \int_0^{mL} w_1(x) \, x \, \mathrm{d}x}{mL}\\
    B_{y1} &= \int_0^{L} w_1(x) \, \mathrm{d}x - \frac{mL\int_0^{mL} w_1(x) \, \mathrm{d}x \,  - \int_0^{mL} w_1(x) \, x \, \mathrm{d}x}{mL} - \frac{\int_{mL}^{L} w_1(x) \, x \, \mathrm{d}x - mL\int_{mL}^{L} w_1(x) \, \mathrm{d}x }{mL} \nonumber \\ 
    C_{y1} &= \frac{\int_{mL}^{L} w_1(x) \, x \, \mathrm{d}x - mL\int_{mL}^{L} w_1(x) \, \mathrm{d}x }{mL} \nonumber \\ 
    A_{y2} &= A_{y1} \pm \frac{RmL}{2} \nonumber \\
    B_{y2} &= B_{y1} \pm \nonumber \bigg(RL - \frac{RmL}{2} - \frac{R(L-mL)^2}{2mL}\bigg)\\ 
    C_{y2} &= C_{y1} \pm \frac{(L-mL)^2R}{2mL} \nonumber \\ 
\end{align}
which allows us to compute:
\begin{align}
    \bigg|A_{y1} - A_{y2}\bigg| &= \bigg|\frac{RmL}{2}\bigg| \\
    \bigg|B_{y1} - B_{y2}\bigg| &= \bigg|RL - \frac{RmL}{2} - \frac{R(L-mL)^2}{2mL}\bigg| \nonumber \\
    \bigg|C_{y1} - C_{y2}\bigg| &= \bigg|\frac{(L-mL)^2R}{2mL}\bigg| \nonumber \\
\end{align}
which can be manipulated, following the example in eqn. \ref{eqn:S_solve}, as:
\begin{align}
    S &= \max\bigg( \bigg|\frac{RmL}{2}\bigg|,\, \bigg|RL - \frac{RmL}{2} - \frac{R(L-mL)^2}{2mL}\bigg|,\, \bigg|\frac{(L-mL)^2R}{2mL}\bigg| \bigg)\\
    S &=
    \begin{cases}
        \big|\frac{(L-mL)^2R}{2mL}\big|,& 0 < m < 1 - \frac{\sqrt{12}}{6}\\
        \big|RL - \frac{RmL}{2} - \frac{R(L-mL)^2}{2mL}\big|,& 1 - \frac{\sqrt{12}}{6} \leq m \leq \frac{1}{2}\\
    \end{cases}
\end{align}
which allows us to determine:
\begin{equation}
    R = \frac{2SmL}{(L-mL)^2}
    \label{eqn:R_3ssc}
\end{equation}
where $0<m \leq \frac{1}{3}$. As a brief note, we define $0<m \leq \frac{1}{3}$ for the general case of $3$ supports in order to accommodate all supports within total length $L$ while simultaneously allowing realizations that place supports at any location throughout the domain. Equation \ref{eqn:R_3ssc} will hold for $0 < m < 1 - \frac{\sqrt{12}}{6}$.

\begin{figure}[h!]
	\centering
	\includegraphics[width=.4\textwidth]{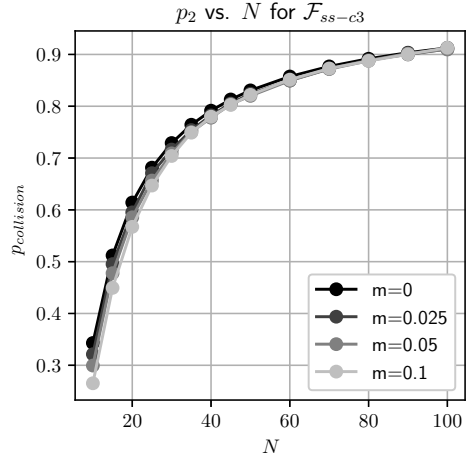}
	\caption{\revs{Simulated probability of a collision ($p_{\mathrm{collision}}$ with respect to the discretization of the input load ($N$) for different values of $m$. For $m=0$ the curve plotted matches eqn. \ref{eqn:p2}.}}
	\label{fig:fig4_apx}
\end{figure}

After we have shown that we can compute $R$ for $p_1=1$, we \nr{then need to show that for some value of $c$ greater than $1$, $p_1$ will be greater than $p_2$. To do this, we} 
conceptualize a worst case example of $w_1(x)$ and $w_2(x)$ where substantially different loads will lead to hash collisions by considering the two loads illustrated in Fig. \ref{fig:fig3_apx}\revs{c}. 
Here, $w_1(x) = \mathbf{0}$ and we can define $w_2(x)$ as a piecewise function:

\begin{align}
w_2(x)& = 
    \begin{cases}
       0,&  x < \frac{tL}{N},\\
       -cR/2,& \frac{tL}{N} \leq x < \frac{(t+1)L}{N},\\
       cR,& \frac{(t+1)L}{N} \leq x < \frac{(t+2)L}{N},\\
       -cR/2,& \frac{(t+2)L}{N} \leq x < \frac{(t+3)L}{N},\\
       0,& \frac{(t+3)L}{N} \leq x\\
    \end{cases} 
\end{align}

where $t$ is an integer, and $L/N$ with $N\geq3$ represents the discretization of the load, \revs{and $c>1$ following the definition introduced with eqn. \ref{eqn:lhs}}.
\nr{In this case, a hash collision will occur when either: (1) $c$ is small enough that $w_2(x)$ will always lead to a change in support force $<S$ regardless of where the support positions are located, or (2) when the support positions defined by distances $l_a$, $l_b$, and $l_c$ (see Fig. \ref{fig:fig2_apx}b-i) all fall outside the range $[tL/N, (t+3)L/N]$.} 
To satisfy the conditions for locality sensitive hashing laid out in eqn. \ref{eqn:lhs}-\ref{eqn:lhs_v2}, we need to show that \nr{arbitrarily far apart functions (i.e., large $c$ and thus large distance between functions $d(w_1, w_2) = cR$)} experience a hash collision with $p_2 < p_1$. In other words, we need to determine \nr{if there is an} upper bound on $p_2$ \nr{as values of $c$ become arbitrarily large}. To do this, we consider scenario (2) and
compute $p_2$ as the probability that \revs{distances $l_a$, $l_b$, and $l_c$ all fall outside the range $[tL/N, (t+3)L/N]$} as a function of our discretization $N$ and our minimum distance between support $m$. In Fig. \ref{fig:fig3_apx}\revs{c}, we plot $p_2$ vs. $N$ for multiple values of $m$. Note that for $m\to0$, we can readily compute:
\begin{equation}
    p_2 = \bigg[\frac{N-3}{N}\bigg]^3
    \label{eqn:p2}
\end{equation}
as the probability that all three supports will be placed outside of the $tL/N \leq x \leq (t+3)L/N$ zone if their locations are randomly generated. From Fig. \revs{\ref{fig:fig4_apx}}, we also see that higher values of $N$ lead to higher values of $p_2$.  
\revs{However, it is clear that even for this worst case of comparison points with large $c$, $p_2 < p_1$ and thus the system is ($R$, $cR$, $p_1$, $p_2$)-\textit{sensitive}.} 
Notably, this simplest case example paves the way for the investigation of more complex mechanical systems as ($R$, $cR$, $p_1$, $p_2$)-\textit{sensitive}.

\section{Example Problem Additional Details}
\label{sec:apx_example}

In Section \ref{sec:example_problem}, we define the example problem that will lead to the results proposed in Section \ref{sec:res_disc}. Here we provide additional information to ensure that our problem definition is clear.

\subsection{Simply Supported Ensemble}
\label{apx:ens}

\begin{figure}[h!]
	\centering
	\includegraphics[width=\textwidth]{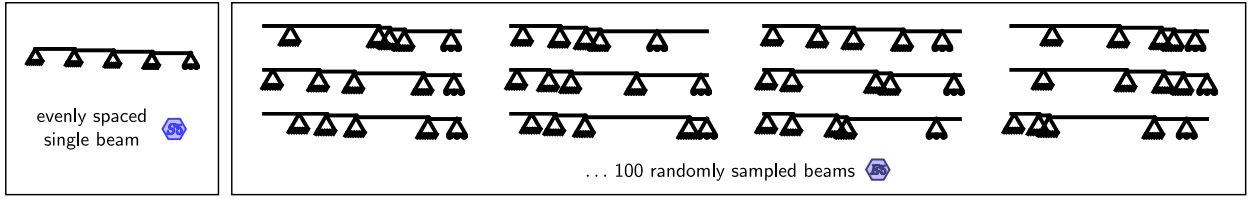}
	\caption{Explicit illustration of the difference between simply supported ``\textbf{S}'' and simply supported ensemble ``\textbf{E}'' mechanical systems for the case with $5$ supports.}
	\label{fig:fig5_apx}
\end{figure}

In Fig. \ref{fig:fig3}b-d, we illustrate the mechanical systems that we explore as hash functions. Here, in Fig. \ref{fig:fig5_apx}, we explicitly illustrate what we mean by an ``ensemble'' of simply supported beams. Namely, each ensemble contains $100$ simply supported beams with randomly generated support locations. Each one of these devices may lead to a different load class prediction, and the final prediction of the ensemble is the hard voting based outcome of combining all $100$ of these predictions. In hard voting, the class labels with the highest frequency from the ensemble predictions becomes the final prediction.

\subsection{Applied Load Categories}
\label{apx:alc}

The applied loads introduce in Section \ref{sec:problem_loads} are described in more detail as follows:
\begin{itemize}
\item \underline{Class 1}, constant load ($c=1/L$):
\begin{equation}
w_1(x) = -c
\end{equation}
\item \underline{Class 2}, piecewise linear load ($c=2/L$):
\begin{align}
w_2(x)& = 
    \begin{cases}
       -c + \frac{2c}{L}x,&  0 \leq x < \frac{L}{2},\\
       (\frac{L}{2} - x)\frac{2c}{L},& \frac{L}{2} \leq x \leq L\\
    \end{cases} 
\end{align}
\item \underline{Class 3}, piecewise linear load ($c=2/L$):
\begin{align}
w_3(x)& = 
    \begin{cases}
       -\frac{2c}{L}x,&  0 \leq x < \frac{L}{2},\\
       \frac{2c}{L}x - 2c,& \frac{L}{2} \leq x \leq L\\
    \end{cases} 
\end{align}
\item \underline{Class 4}, linear load ($c=2/L$):
\begin{equation}
\ w_4(x) = \frac{c}{L}x - c
\end{equation}
\item \underline{Class 5}, linear load ($c=2/L$):
\begin{equation}
\ w_5(x) = -\frac{c}{L}x
\end{equation}
\item \underline{Class 6}, sine wave with wave number $k=0.5$ and offset $\varphi=0$:
\begin{equation}
\label{eqn:sin}
\ w_6(x) = - \frac{\pi}{2L}\bigg| \sin\bigg( \frac{k}{2 \pi L}x - 2 \pi \varphi \bigg) \bigg|
\end{equation}
\item \underline{Class 7}, sine wave with wave number $k=1.0$ and offset $\varphi=0$, see eqn. \ref{eqn:sin}.
\item \underline{Class 8}, sine wave with wave number $k=1.0$ and offset $\varphi=0.25$, see eqn. \ref{eqn:sin}.
\item \underline{Class 9}, sine wave with wave number $k=1.5$ and offset $\varphi=0$, see eqn. \ref{eqn:sin}.
\item \underline{Class 10}, sine wave with wave number $k=1.5$ and offset $\varphi=0.25$, see eqn. \ref{eqn:sin}.
\item \underline{Class 11}, sine wave with wave number $k=2.0$ and offset $\varphi=0$, see eqn. \ref{eqn:sin}.
\item \underline{Class 12}, negative kernel density estimate (kde) based on $n=2$ points $p_i$ with uniform random location on the x-axis. The negative kde for a Gaussian kernel $\rho_K(x)$ with  bandwidth $h=0.1L$ is written as:
\begin{equation}
\label{eqn:kde}
\ w_{12}(x) = -\frac{1}{nh} \sum\limits_{i=1}^{i=n} \exp \bigg( \frac{-(x-p_i)^2}{2h^2} \bigg)
\end{equation}
\item \underline{Class 13}, negative kernel density estimate (kde) based on $n=2$ points $p_i$ with uniform random location on the x-axis, see eqn. \ref{eqn:kde}.
\item \underline{Class 14}, negative kernel density estimate (kde) based on $n=2$ points $p_i$ with uniform random location on the x-axis, see eqn. \ref{eqn:kde}.
\item \underline{Class 15}, negative kernel density estimate (kde) based on $n=5$ points $p_i$ with uniform random location on the x-axis, see eqn. \ref{eqn:kde}.
\item \underline{Class 16}, negative kernel density estimate (kde) based on $n=5$ points $p_i$ with uniform random location on the x-axis, see eqn. \ref{eqn:kde}.
\item \underline{Class 17}, negative kernel density estimate (kde) based on $n=5$ points $p_i$ with uniform random location on the x-axis, see eqn. \ref{eqn:kde}.
\item \underline{Class 18}, negative kernel density estimate (kde) based on $n=25$ points $p_i$ with uniform random location on the x-axis, see eqn. \ref{eqn:kde}.
\item \underline{Class 19}, negative kernel density estimate (kde) based on $n=25$ points $p_i$ with uniform random location on the x-axis, see eqn. \ref{eqn:kde}.
\item \underline{Class 20}, negative kernel density estimate (kde) based on $n=25$ points $p_i$ with uniform random location on the x-axis, see eqn. \ref{eqn:kde}.
\end{itemize}
Each load is set up so that the area under the curve is equal to $1$.
There are $20$ examples for each of the $20$ classes of loads. 
The $20$ examples are all differentiated from each other through the addition of Perlin noise with randomly selected initial seed and integer octave in range $[2-10]$. 
Details for accessing the code to exactly reproduce these loads including the randomly generated Perlin noise are given in Section \ref{sec:additional_info}.

\section{Results Additional Details}
\label{sec:res_apx}

\begin{figure}[h!]
	\centering
	\includegraphics[width=0.6\textwidth]{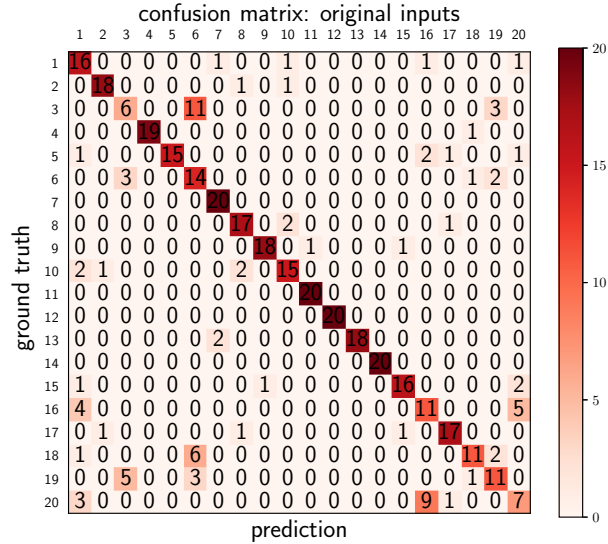}
	\caption{Confusion matrix based on the original input loads. Note that this confusion matrix corresponds to a classification accuracy of $0.77$ across $20$ classes.}
	\label{fig:fig6_apx}
\end{figure}

\begin{figure}[p!]
	\centering
	\includegraphics[width=\textwidth,trim={0mm 40mm 0mm 0mm},clip]{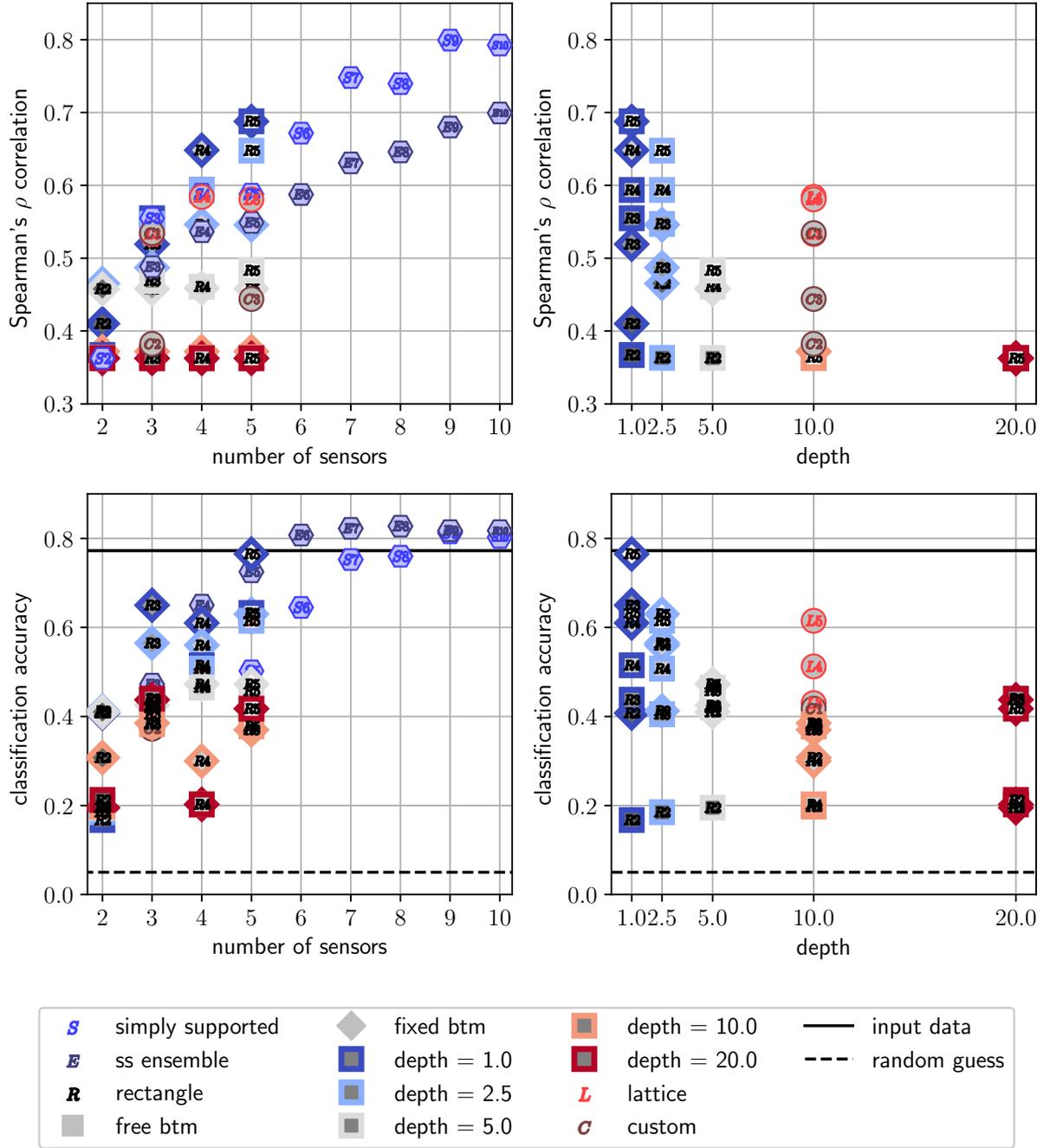}
	\caption{\revs{As a supplement to Fig. \ref{fig:fig5}, we provide Spearman's $\rho$ and classification accuracy plotted with respect to both number of sensors and domain depth.}}
	\label{fig:fig12}
\end{figure}

\revs{This Appendix contains additional supporting results to supplement the information presented in Section \ref{sec:res_disc}. In Section \ref{sec:res_disc}, we compare the classification accuracy of our mechanical systems to both random guessing $accuracy=0.05$ and direct analysis of the original input data $accuracy=0.77$.
Here, in Fig. \ref{fig:fig6_apx}, we show the confusion matrix for nearest neighbor load classification based on the original input data. The purpose of showing this graphic is to demonstrate that we have chosen a challenging suite of applied loads that are non-trivial to distinguish. Next, for completeness and as an alternative view of the relationship between Spearman's $\rho$, classification accuracy, and mechanical domain properties, we provide Fig. \ref{fig:fig12} as a supplement to Fig. \ref{fig:fig5} and Fig. \ref{fig:fig6}. Finally, we provide Table \ref{tab:tab1} which contains Spearman's $\rho$ and classification accuracy for every system investigated in this study. These data are directly visualized in Fig. \ref{fig:fig5}, Fig. \ref{fig:fig6}, and Fig. \ref{fig:fig12}.}

\begin{table}[h]
{\small
\centering
\begin{tabular}{@{}ccccc@{}}
\toprule
system type      & number of sensors $ns$ & domain depth $d$ & Spearman's $\rho$ & classification accuracy \\ \midrule
simply supported & 2                      & n/a              & 0.36              & 0.17                    \\
simply supported & 3                      & n/a              & 0.55              & 0.42                    \\
simply supported & 4                      & n/a              & 0.59              & 0.5                     \\
simply supported & 5                      & n/a              & 0.59              & 0.5                     \\
simply supported & 6                      & n/a              & 0.67              & 0.65                    \\
simply supported & 7                      & n/a              & 0.75              & 0.75                    \\
simply supported & 8                      & n/a              & 0.74              & 0.76                    \\
simply supported & 9                      & n/a              & 0.8               & 0.81                    \\
simply supported & 10                     & n/a              & 0.79              & 0.8                     \\
ss ensemble      & 3                      & n/a              & 0.49              & 0.47                    \\
ss ensemble      & 4                      & n/a              & 0.54              & 0.65                    \\
ss ensemble      & 5                      & n/a              & 0.55              & 0.72                    \\
ss ensemble      & 6                      & n/a              & 0.59              & 0.81                    \\
ss ensemble      & 7                      & n/a              & 0.63              & 0.82                    \\
ss ensemble      & 8                      & n/a              & 0.65              & 0.83                    \\
ss ensemble      & 9                      & n/a              & 0.68              & 0.82                    \\
ss ensemble      & 10                     & n/a              & 0.7               & 0.82                    \\
rect fixed btm   & 2                      & 1                & 0.41              & 0.41                    \\
rect fixed btm   & 3                      & 1                & 0.52              & 0.65                    \\
rect fixed btm   & 4                      & 1                & 0.65              & 0.61                    \\
rect fixed btm   & 5                      & 1                & 0.69              & 0.77                    \\
rect fixed btm   & 2                      & 2.5              & 0.47              & 0.41                    \\
rect fixed btm   & 3                      & 2.5              & 0.49              & 0.56                    \\
rect fixed btm   & 4                      & 2.5              & 0.55              & 0.56                    \\
rect fixed btm   & 5                      & 2.5              & 0.55              & 0.63                    \\
rect fixed btm   & 2                      & 5                & 0.46              & 0.41                    \\
rect fixed btm   & 3                      & 5                & 0.46              & 0.42                    \\
rect fixed btm   & 4                      & 5                & 0.46              & 0.47                    \\
rect fixed btm   & 5                      & 5                & 0.46              & 0.47                    \\
rect fixed btm   & 2                      & 10               & 0.37              & 0.31                    \\
rect fixed btm   & 3                      & 10               & 0.37              & 0.39                    \\
rect fixed btm   & 4                      & 10               & 0.37              & 0.3                     \\
rect fixed btm   & 5                      & 10               & 0.37              & 0.37                    \\
rect fixed btm   & 2                      & 20               & 0.36              & 0.2                     \\
rect fixed btm   & 3                      & 20               & 0.36              & 0.44                    \\
rect fixed btm   & 4                      & 20               & 0.36              & 0.2                     \\
rect fixed btm   & 5                      & 20               & 0.36              & 0.42                    \\
rect             & 2                      & 1                & 0.37              & 0.17                    \\
rect             & 3                      & 1                & 0.55              & 0.44                    \\
rect             & 4                      & 1                & 0.59              & 0.52                    \\
rect             & 5                      & 1                & 0.69              & 0.63                    \\
rect             & 2                      & 2.5              & 0.36              & 0.18                    \\
rect             & 3                      & 2.5              & 0.55              & 0.41                    \\
rect             & 4                      & 2.5              & 0.59              & 0.51                    \\
rect             & 5                      & 2.5              & 0.65              & 0.61                    \\
rect             & 2                      & 5                & 0.36              & 0.2                     \\
rect             & 3                      & 5                & 0.47              & 0.42                    \\
rect             & 4                      & 5                & 0.46              & 0.47                    \\
rect             & 5                      & 5                & 0.48              & 0.46                    \\
rect             & 2                      & 10               & 0.36              & 0.2                     \\
rect             & 3                      & 10               & 0.36              & 0.38                    \\
rect             & 4                      & 10               & 0.36              & 0.2                     \\
rect             & 5                      & 10               & 0.36              & 0.38                    \\
rect             & 2                      & 20               & 0.36              & 0.21                    \\
rect             & 3                      & 20               & 0.36              & 0.44                    \\
rect             & 4                      & 20               & 0.36              & 0.2                     \\
rect             & 5                      & 20               & 0.36              & 0.42                    \\
lattice          & 3                      & 10               & 0.53              & 0.43                    \\
lattice          & 4                      & 10               & 0.58              & 0.51                    \\
lattice          & 5                      & 10               & 0.58              & 0.61                    \\
custom 1         & 3                      & 10               & 0.53              & 0.42                    \\
custom 2         & 3                      & 10               & 0.38              & 0.37                    \\
custom 3         & 5                      & 10               & 0.44              & 0.37                    \\ \bottomrule
\end{tabular}
\caption{\label{tab:tab1} Summary of results, supporting information for Fig. \ref{fig:fig5}.}}
\end{table}

\FloatBarrier
\newpage

\bibliographystyle{unsrt}

\end{document}